\documentclass[12pt,letterpaper,oneside,english]{iopart}
\usepackage{times}
\usepackage[T1]{fontenc}
\usepackage[latin1]{inputenc}
\usepackage{babel}
\usepackage{graphicx}
\usepackage{amssymb}

\newcommand{\nd}{\noindent}
\newcommand{\be}{\begin{equation}}
\newcommand{\ee}{\end{equation}}
\newcommand{\ben}{\begin{eqnarray}}
\newcommand{\een}{\end{eqnarray}}

\newtheorem{theorem}{Theorem}

\newtheorem{definition}[theorem]{Definition}

\begin{document}

\title[Elliptical invariance of distributions of the power type: the stability and extensivity issues] {Elliptical invariance of distributions of the power type: the stability and extensivity issues}

\author{C. Vignat \dag, A. Plastino \ddag}
\address{\dag Laboratoire de Production Microtechnique, E.P.F.L., Switzerland}
\address{\ddag IFLP-Exact Sciences Faculty, National University La Plata and
Argentina's CONICET, C. C. 727, 1900 La Plata, Argentina}
\ead{vignat@epfl.ch}

\begin{abstract}
In this paper we delve into some important properties of probability distributions of the  power type in order to provide some
answers to questions recently raised in the literature. More
precisely, we focus on the properties of maximizers of generalized
information measures and give results about their stability under
addition-composition processes.
\end{abstract}
\submitto{JSTAT} 


\maketitle

\section{Introduction}
\nd Non-logarithmic information measures have become vary fashionable nowadays,
with multiple applications to different scientific disciplines
(see, for instance, \cite{euro} and references therein). They were
introduced in the cybernetic-information communities by
Harvda-Charvat \cite{l1} in 1967 and Vadja  \cite{l2} in 1968, and
rediscovered by Daroczy in 1970  \cite{l3}  with several echoes mostly in
the field of image processing: see \cite{okamoto} for a historic summary and the
pertinent references. In astronomy, physics,
economics, biology etc..., these non-logarithmic information
measures are often used under the form of the $q-$entropies as introduced by Tsallis since 1988 \cite{t0}.
\vskip 1mm \nd These entropies are maximized by power-type
distributions. The properties of both discrete and
continuous power-type distributions have been carefully reviewed recently in Ref.
\cite{Tsallis} in what respects to \begin{enumerate} \item their
behavior by convolution and \item
 their relationships with stable L\'{e}vy distributions. \end{enumerate}  
In this paper, we wish to focus attention more closely on further
properties of these distributions, and answer some open questions as raised in \cite{Tsallis}; this way, we hope,  in the wake of Refs.
\cite{berg,johnson,next2003}, to positively contribute to a more
complete understanding  of the ensuing
theoretical context.

\section{Definitions and Notations}

\nd In what follows we consider some probability
density $f_{X} \, (X \in \mathbb{R}^N)$ that maximize a
generalized entropy, either of the Harvda-Charvat-R\'{e}nyi type
\begin{equation}
H_{q}\left(X\right)=\frac{1}{1-q}
\log \left(\int_{\mathbb{R}^{n}}f_{X}^{q}\left(X\right)dX\right).\label{eq:Hq}\end{equation}
or of the Tsallis type
\begin{equation}
S_{q}\left(X\right)=\frac{1}{q-1}
\left(1-\int_{\mathbb{R}^{n}}f_{X}^{q}\left(X\right)dX\right).\label{eq:Sq}\end{equation}
where $q$ is a real parameter (called ''nonextensivity
parameter" in \cite{euro}). As $H_{q}$ can be expressed as an increasing function of $S_{q}$, both entropies have the same maximizers. As a consequence, all results expressed in this paper hold for both types of entropies, except in Section \ref{sectionext} that deals with a special property of $S_{q}$.
To each density $f_{X}$, we associate its so-called escort
distribution  \cite{beck} defined as \[
F_{X}\left(X\right)=\frac{f_{X}^{q}
\left(X\right)}{\int_{\mathbb{R}^{n}}f_{X}^{q}\left(X\right)dX}.\]
Note that the dependence of $F_{X}$ on $q$ is not explicitly
stated for notational simplicity.

\subsection{Power-law distributions as entropy-maximizers}

The following theorem  generalizes to the
$n-$variate case the characterization given in  Ref. \cite[Eq.
(42)]{Tsallis} for the maximum entropy distributions with fixed
$q-$covariance.

\begin{theorem}
Under the q-covariance constraint \[ \int
XX^{T}F_{X}\left(X\right)dX=K\] (where the $q-$covariance matrix
$K$ is symmetric definite positive) and the normalization
constraint $\int f_{X}=1$, the power-law entropy
(\ref{eq:Hq}) or (\ref{eq:Sq}) has a single maximizer equal to:

$\bullet$ if $1<q<\frac{n+2}{n}$\newline \nd \fbox{\parbox{4in}{
\begin{equation}  f_{X}\left(X\right)=
A_{q}\left(1+X^{T}\Lambda^{-1}X\right)^{\frac{1}{1-q}}\label{eq:Student-t}\end{equation}}}
\newline \nd with

\[
A_{q}=\frac{\Gamma\left(\frac{1}{q-1}\right)}{\Gamma\left(\frac{1}{q-1}-\frac{n}{2}\right)\vert\pi\Lambda\vert^{1/2}}, \Lambda=mK, m=\frac{2}{q-1}-n.\]

$\bullet$ if $q<1$\newline \nd \fbox{\parbox{4in}{\begin{equation}
f_{X}\left(X\right)=A_{q}\left(1-X^{T}\Sigma^{-1}X\right)_{+}^{\frac{1}{1-q}}
\label{eq:Student-r}\end{equation}}}\newline \nd  with\[
A_{q}=\frac{\Gamma\left(\frac{2-q}{1-q}+\frac{n}{2}\right)}{\Gamma\left(\frac{2-q}{1-q}\right)\vert\pi\Sigma\vert^{1/2}}, \Sigma=pK, p=\frac{2}{1-q}+n\]
and with notation $\left(x\right)_{+}=\max\left(0,x\right)$.
\end{theorem}
\nd In the case $n=1$ we recover the results of \cite[eq.
(42)]{Tsallis}, namely

\begin{itemize}
\item if $1<q<3$
\end{itemize}
\begin{equation}
f_{X}\left(x\right)=\frac{\Gamma\left(\frac{1}{q-1}\right)\sqrt{q-1}}{\Gamma\left(\frac{3-q}{2\left(q-1\right)}\right)\sqrt{3-q}\sigma\sqrt{\pi}}\left(1+\left(\frac{q-1}{3-q}\right)\frac{x^{2}}{\sigma^{2}}\right)^{\frac{1}{1-q}}\label{eq:Student-t_1d}\end{equation}

\begin{itemize}
\item if $q<1$\begin{equation}
f_{X}\left(x\right)=
\frac{\Gamma\left(\frac{5-3q}{2\left(1-q\right)}\right)\sqrt{1-q}}
{\Gamma\left(\frac{2-q}{1-q}\right)\sqrt{3-q}\sigma\sqrt{\pi}}
\left(1-\left(\frac{1-q}{3-q}\right)
\frac{x^{2}}{\sigma^{2}}\right)_{+}^{\frac{1}{1-q}}.\label{eq:Student-r_1d}\end{equation}

\end{itemize}
Note the existence of a minor typo in \cite{Tsallis} for the
definition of $A_{q}$ in the case $q>1$ (replace $2/(1-q)$ by
$2\left(1-q\right)$). For the correct expression see also
\cite{Prato}.

\subsection{Student-t and Student-r distributions}

In statistics, distribution (\ref{eq:Student-t}) is called an $n-$variate
Student-t with $m$ degrees of freedom and $q-$covariance matrix
$K$: it will be denoted as $\mathcal{T}\left(m,n,K\right)$ in the
following. We notice that its nonextensivity parameter $q$ is
linked to the dimension $n$ and the number of degrees of freedom
$m$ by \[ q=\frac{m+n+2}{m+n}.\] Moreover, convergence of both integrals  $\int f_{X}\left(X\right)dX$ and $\int
XX^{T}F_{X}\left(X\right)dX$ requires the same condition, namely
$q<\frac{n+2}{n},$ or equivalently $m>0.$ In the next section,
we will endow parameter $m$ with a meaning.


\nd Accordingly, distribution (\ref{eq:Student-r}) is an
$n-$variate Student-r with $p$ degrees of freedom and
$q-$covariance matrix $K$: it will be denoted as
$\mathcal{R}\left(p,n,K\right).$ We remark that its nonextensivity
parameter $\,q$ is linked to parameter $p$ and dimension $n$ as\[
q=\frac{p-n-2}{p-n}.\]

\subsection{Stochastic representations}

Beck and Cohen \cite{beck} have recently introduced in the
literature an interesting statistical concept,  baptized with the
name {\it superstatistics}, that links different types of
probability densities. In this vein, our  distributions above can
be shown to correspond to {\it multivariate Gaussian densities}
whose covariance matrix fluctuates according to a certain law, as
detailed in the two following theorems.

\begin{theorem}
\label{thm:stochastic}If $X$ follows a $\mathcal{T}\left(m,n,K\right)$
distribution then a stochastic representation of $X$ writes %
\footnote{in the following, sign $\sim$ means {}``is distributed as''%
}\begin{equation}
X\sim\frac{\Lambda^{1/2}G}{a}\label{eq:stochastic_t}\end{equation}
where $G$ is an $n-$variate Gaussian vector with unit covariance
matrix, $a$ is a random variable independent of $G$ that follows
a $\chi$ distribution%
\footnote{a chi distribution is $f_{a}(a) = \frac{2^{1-m/2}}{\Gamma(\frac{m}{2})} a^{m-1} \exp(-a^{2}/2)$; chi distributions are restricted to integer degrees of freedom. If
$m\notin\mathbb{N}$ then the $\chi$ distribution should be extended
to the distribution of the square-root of a gamma random variable
with shape parameter equal to $2m$. For the sake of simplicity, we
will speak of $\chi$ distribution in this case too.%
} with number of degrees of freedom $m=\frac{2}{q-1}-n$ and $\Lambda=mK.$
\end{theorem}
\nd A remarkable fact deserves here emphasizing upon:  this
approach can be {\it extended} to the case when $q<1$, with a
noticeable difference. This extension is based on the following
duality result.

\begin{theorem}
if $X\sim\mathcal{T}\left(m,n,K\right)$ and $\Lambda=mK$ then random
vector $Y$ defined as\begin{equation}
Y=\frac{X}{\sqrt{1+X^{T}\Lambda^{-1}X}}\label{eq:transf}\end{equation}
is such that $Y\sim\mathcal{R}\left(p,n,K\right)$ with\[
p=m+n-2.\]
If $q$ and $q'$ denote the respective nonextensivity indices of
$X$ and $Y,$ then
\be
\label{qpq}
\frac{1}{1-q'}=\frac{1}{q-1}-\frac{n}{2}-1.
\ee

\end{theorem}
In Figure \ref{fig1} below,
values of $q'$ as a function of $q$ are plotted
for $n=1, 2, 5$ and $10$ (right to left). We remark that
transformation (\ref{eq:transf}) induces a one-to-one relationship
between $q\in\left[1,\frac{n+4}{n+2}\right[$ and
$q'\in\left]-\infty,1\right]$ and has the Gaussian distribution
($q=q'=1$) as fixed point.

\begin{figure}[h] 
   \centering
   \includegraphics[scale=0.4]{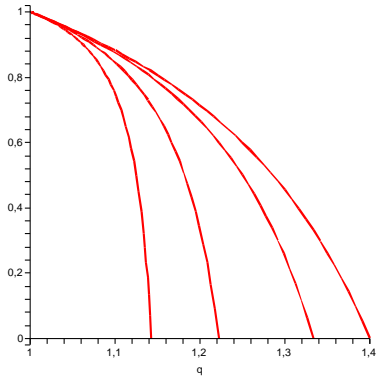} 
   \caption{$q'$ as a function of $q$ as in (\ref{qpq}) for $n=1, 2, 5$ and $10$ (right to left)}
   \label{fig1}
\end{figure}

\nd An important consequence of the above is the following dual
result of theorem (\ref{thm:stochastic}).

\begin{theorem}
If $Y$ follows a $\mathcal{R}\left(p,n,K\right)$ distribution then
a stochastic representation of $Y$ writes\begin{equation}
Y\sim\frac{\Sigma^{1/2}G}{\sqrt{G^{T}G+b^{2}}}\label{eq:stochastic_r}\end{equation}
where $G$ is an $n-$variate Gaussian vector with unit covariance
matrix, where $\Sigma=\left(p-n+2\right)K$ and $b$ is a random variable
independent of $G$ that follows a $\chi$ distribution with $p-n+2$
degrees of freedom.
\end{theorem}
\nd Here, the important difference, as compared to the case $q>1$,
is to be found in the fact that  the fluctuations, represented by
the denominator of (\ref{eq:stochastic_r}), are now
\textit{dependent} of the values of the Gaussian system through
the presence of term $G^{T}G$.

\subsection{Covariance matrices}

The covariance matrices $R=EXX^{T}$ of both distributions are related to their
$q-$covariance matrices as follows.

\begin{theorem}
Distribution $\mathcal{T}\left(m,n,K\right)$ has covariance matrix
\be
\label{RK1}
R=\frac{m}{m-2}K
\ee
provided $m>2,$ that is $q<\frac{n+4}{n+2}$. For example, a finite
covariance matrix exists in the case $n=1$ only if $1<q<\frac{5}{3}.$
\end{theorem}
\vskip 1mm {\sf Proof} \vskip 1mm \nd Using the stochastic
representation (\ref{eq:stochastic_t}), we deduce\[
EXX^{T}=EGG^{T}E\frac{1}{a^{2}}, \] with $Ea^{-2}=\frac{1}{m-2}$
and $EGG^{T}=\Lambda=mK  \,\, \Box$ \vskip 1mm
\begin{theorem}
Distribution $\mathcal{R}\left(p,n,K\right)$ has covariance matrix
\be
\label{RK2}
R=\frac{p-n+2}{p+2}K. 
\ee
\end{theorem}
\vskip 1mm  {\sf Proof} \vskip 1mm \nd The proof uses the polar
factorization property \cite{barthe} of stochastic representation
(\ref{eq:stochastic_r}), namely the fact that
$\frac{G}{\sqrt{G^{T}G+b^{2}}}$ and $\sqrt{G^{T}G+b^{2}}$ are
independent. As a consequence\[
EYY^{T}=\frac{\Sigma^{1/2}EGG^{T}\Sigma^{1/2}}
{E\left(G^{T}G+b^{2}\right)}=\frac{p-n+2}{p+2}K \,\, \Box \]

\vskip 1mm  \nd We note that in the Gaussian case
($p\rightarrow+\infty$ in (\ref{RK2}) or $m\rightarrow+\infty$ in (\ref{RK1})), the
$q-$covariance and the variance matrices coincide.

\subsection{Geometric characterization}

Geometric characterizations of both distributions
(\ref{eq:Student-t}) and (\ref{eq:Student-r}) in terms of
projections of the uniform distribution on the sphere in
$\mathbb{R}^{n}$ are detailed in \cite{next2005}.
According to the stochastic representation
(\ref{eq:stochastic_r}), $\Sigma^{-1/2}Y$ can be interpreted, if $p\in\mathbb{N}$, as the marginal vector of a $\left(p+2\right)-$variate random vector uniformly
distributed on the sphere in $\mathbb{R}^{p+2}.$
A link between this observation and the role of extended information measures in the microcanonical framework can be found in \cite{next2005}.

\section{The stability issue}

\nd As noted in \cite{Tsallis}, distributions (\ref{eq:Student-t})
and (\ref{eq:Student-r}) are not stable by convolution since they
do not belong to the L\'{e}vy class: the sum of two independent random
variables following either distribution
(\ref{eq:Student-t}) or distribution (\ref{eq:Student-r}) does
not follow any of these distributions again, as opposed to the
Maxwellian-Gaussian case. It is then suggested in \cite{Tsallis}
 that, in order to recover the original distribution after
summation, a certain kind of dependence should be
introduced between the components of the sum.
\newline \nd
It is the aim of the next paragraph to show that such  dependence can be accurately characterized
in the case of power-law distributions.

\subsection{A first example: case $q>1$}

\nd Let us assume, for instance, that $q>1$ and choose $X$ to be a
random vector of dimension $n$ distributed according to
(\ref{eq:Student-t}). We extract from it two scalar
components, say $X_{1}$ and $X_{2}$; according to
(\ref{eq:stochastic_t}), these two components can be expressed as
\begin{equation}
X_{1}\sim\frac{\left(\Lambda^{1/2}G\right)_{1}}{a}, X_{2}\sim\frac{\left(\Lambda^{1/2}G\right)_{2}}{a}\label{eq:extracted1d}\end{equation}
where $\left(.\right)_{1}$ denotes the first vector component.

\vskip 3mm {\sf Distribution of the components} \vskip 2mm

 \nd  We first remark from stochastic representation
(\ref{eq:extracted1d}) that $X_{1}$ and $X_{2}$ are again
distributed according to a Student-t distribution with dimension
$n=1$; moreover, the extraction of components keeps the
fluctuation variable $a$ unchanged, so that both $X_{1}$ and
$X_{2}$ have unchanged number of degrees of freedom
$m'=m=\frac{2}{q-1}-n$. Both have thus a new nonextensivity
parameter $q'$ that verifies\[ \frac{2}{q'-1}-1=\frac{2}{q-1}-n\]
or equivalently
\be
\label{qpq2}
q'=1+\frac{2\left(q-1\right)}{2+\left(1-q\right)\left(n-1\right)}.
\ee

\begin{figure}[h] 
   \centering
   \includegraphics[scale=0.4]{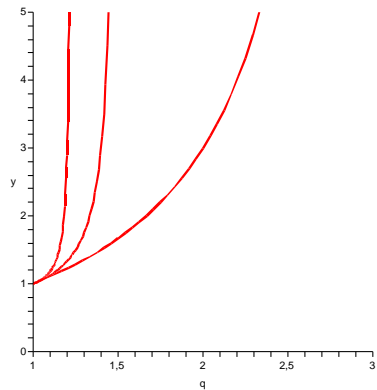} 
   \caption{$q'$ as a function of $q$ as in (\ref{qpq2}) for $n=2, 5$ and $10$  (right to left)}
   \label{fig2}
\end{figure}

\nd Moreover, it is easy to check that their respective $q-$variances
are $K_{11}$ and $K_{22}$, the two first diagonal entries of
$q-$covariance matrix $K$. The three curves in Figure \ref{fig2} represent
$q'$ as a function of $q$ for
$n=2, 5$ and $10$  (from right to left).

\nd We note that

\begin{itemize}
\item $q=1\Rightarrow q'=1$, since any component of a Gaussian vector is
Gaussian
\item the nonextensivity parameter $q'$ of a single component is larger
than the nonextensivity parameter $q$ of the system it is extracted
from
\item moreover, $q'$ is all the larger since the dimension $n$ is
large.
\end{itemize}

\vskip 3mm {\sf Distribution of the convolution} \vskip 2mm \nd
The distribution of a linear combination $Z$ of $X_{1}$ and
$X_{2}$ can be computed as

\ben & Z=\alpha X_{1}+\beta X_{2}
\sim\frac{1}{a}\left(\alpha\left(\Lambda^{1/2}G\right)_{1}
+\beta\left(\Lambda^{1/2}G\right)_{2}\right)  \cr &
\sim\sqrt{m}\sqrt{\alpha^{2}K_{11}+\beta^{2}K_{22}+2\alpha\beta
K_{12}}\frac{G}{a},\een so that $Z$ is again distributed as a
Student-t distribution with
same parameter $m$ and $q-$variance
$\alpha^{2}K_{11}+\beta^{2}K_{22}+2\alpha\beta K_{12}.$
We underline the fact that stability under convolution originates
from the special type of dependence that exists between the
components $X_{1}$ and $X_{2},$ namely from the fact that they
belong to a same (larger)  system: in more physical terms, $X_{1}$
and $X_{2}$ are components that have experienced the same random
source of fluctuations.

\subsection{A second example: case $q<1$}

\nd We assume now that we extract two components $Y_{1}$ and
$Y_{2}$ from a vector $Y\sim\mathcal{R}\left(p,n,K\right).$
Then a stochastic representation of $Y_{1}$ and $Y_{2}$ is\[
Y_{1} \sim \frac{\left(\Sigma^{1/2}G\right)_{1}}{\sqrt{G^{T}G+b^{2}}}, Y_{2} \sim \frac{\left(\Sigma^{1/2}G\right)_{2}}{\sqrt{G^{T}G+b^{2}}}\]
so that $Y_{1}$ (resp. $Y_{2}$) follows a distribution
$\mathcal{R}\left(p',1,K'\right)$ with $p'=p, K'=K_{11}$ (resp.
$K'=K_{22}$) and its new index of nonextensivity verifies\[
\frac{2}{1-q'}+1=\frac{2}{1-q'}+n\] or
\be
\label{qpq3}
q'=1-\frac{2\left(1-q\right)}{2+\left(n-1\right)\left(1-q\right)}.
\ee
We remark that (\ref{qpq3}) coincides with (\ref{qpq2}) since conservation of degrees of freedom $m$ in the Student-t case and $p$ in the Student-r case is expressed by the same condition.

\nd In Figure \ref{fig3} below, $q'$ is represented as a function of $q$ for
$n=2, 5$ and $10$ (bottom to top).

\begin{figure}[h] 
   \centering
   \includegraphics[scale=0.4]{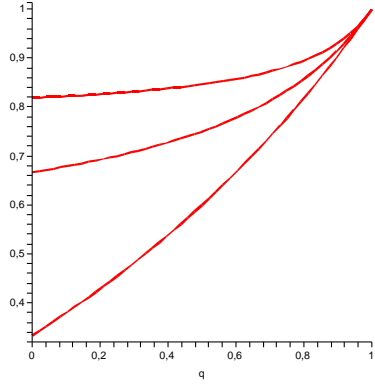} 
   \caption{$q'$ as a function of $q$ as in (\ref{qpq3}) for
$n=2, 5$ and $10$ (bottom to top)}
   \label{fig3}
\end{figure}

\nd The same conclusions as in the case $q>1$ hold, namely:

\begin{itemize}
\item if $q=1$ then $q'=1$ (Gaussian case)
\item the nonextensivity parameter $q'$ of an extracted component is always
larger than the nonextensivity parameter of the original system; it is all the larger since the dimension $n$ is large
\end{itemize}
The distribution of a linear combination can be evaluated as

\ben & Z=\alpha Y_{1}+\beta
Y_{2}\sim\frac{\alpha\left(\Sigma^{1/2}G\right)_{1}
+\beta\left(\Sigma^{1/2}G\right)_{2}}{\sqrt{G^{T}G+b^{2}}} \cr &
\sim\sqrt{p}\sqrt{\alpha^{2}K_{11}+\beta^{2}K_{22}+2\alpha\beta
K_{12}}\frac{G}{\sqrt{G^{T}G+b^{2}}},\een 
so that
$Z\sim\mathcal{R}\left(p,1,\alpha^{2}K_{11}+\beta^{2}K_{22}+2\alpha\beta
K_{12}\right).$

\subsection{Orthogonal invariance}

These results can be generalized using the notion
of elliptical distribution \cite{Chu}.

\begin{definition}
A distribution $f_{X}$ is elliptical (or elliptically invariant)
if it writes\[
f_{X}\left(X\right)=\phi\left(X^{T}C_{X}^{-1}X\right)\]
for some positive definite matrix $C_{X}$ called the characteristic
matrix of $f_{X}$ and some function $\phi$ that may depend of $n.$
\end{definition}
\nd From (\ref{eq:Student-t}) and (\ref{eq:Student-r}), we check
immediately that Student-t and Student-r distributions are elliptically
invariant. This special property can be
justified as follows:
up to application of the mapping $X\rightarrow\Lambda^{1/2}X$
or $Y\rightarrow\Sigma^{1/2}Y$, it may be assumed in
(\ref{eq:Student-t}) and (\ref{eq:Student-r}) that $\Lambda=I_{n}$ or
$\Sigma=I_{n}:$ this special case of elliptical invariance is called
spherical invariance. An equivalent definition of spherical
invariance reads as follows:  for all orthogonal matrices $O,$ the
distribution of $X$ coincides with the distribution of $X$:\[
f_{OX}\left(X\right)=f_{X}\left(X\right).\] Now, the $H_{q}$ or $S_q-$entropy
remains unchanged by orthogonal transformation since, for example
\ben
& S_{q}\left(OX\right)=\frac{1}{q-1}\left(1-\int
f_{OX}^{q}\right) \cr & =\frac{1}{q-1}\left(1-\frac{1}{\vert
O\vert^{q-1}}\int f_{X}^{q}\right)=S_{q}\left(X\right)
\een where we
have used the fact that for any orthogonal matrix, $\vert
O\vert=1.$ Moreover, the constraints under which the $S_q-$entropy
is maximized, that is\[ \int XX^{T}F_{q}\left(X\right)dX=I_{n},\int
f\left(X\right)dX=1\] are themselves spherically invariant as
well. Thus, it is not surprising that the maximizer of $S_{q}$ under
these constraints is spherically invariant - and elliptically
invariant in the more general case $C_{X}\ne I_{n}$.

\subsection{Properties of elliptical distributions and consequences}

The stability property exposed in parts 3.1 and 3.2 appears as a
particular case of the more general property of elliptical
distributions  that we cast here as follows:

\begin{theorem}\cite{Chu}
If $X$ is distributed according to an elliptical distribution\[
f_{X}\left(X\right)=\phi\left(X^{T}C_{X}^{-1}X\right)\]
and if $A$ is a $\left(p\times n\right)$ full-rank matrix with $p\le n$
then $\tilde{X}=AX$ is again elliptically invariant with characteristic
matrix\begin{equation}
C_{\tilde{X}}=AC_{X}A^{T}.\label{eq:AKAT}\end{equation}

\end{theorem}
\nd As a consequence, one can characterize the precise way in
which power-law random vectors behave under linear
transformation as follows.

\vskip 3mm
 {\sf Case of components' extraction}
\vskip 2mm \nd Suppose we extract the $k<n$ first components
$X'=\left(X_{1},\dots,X_{k}\right)$ from a vector of the power-law
type $X\sim\mathcal{T}\left(m,n,K\right)$.  This process
corresponds to applying the matrix \[ A=\left[I_{k\times
k} \vdots O_{\left(n-k\right)\times k}\right]\] to vector $X$, and
we conclude that $X'=AX\sim\mathcal{T}\left(m',k,K'\right)$ where
$K'=AKA^{T}$ coincides with the principal $\left(k\times k\right)$
block of $K$ and $m'=m$, corresponding to a new index of
extensivity\begin{equation}
q'=1+\frac{2\left(q-1\right)}{2-\left(n-k\right)\left(q-1\right)}>1.\label{eq:qp_q_t_case}\end{equation}

\nd For a power law  vector $Y\sim\mathcal{R}\left(p,n,K\right)$, as remarked in part 3.2, conservation of the number $p$ of degrees of freedom yields the same condition as conservation of the number $m$ of degrees of freedom in the Student-t case, that is 

\[ \frac{2}{1-q'}+k=\frac{2}{1-q}+n\] 
or
\begin{equation}
q'=1-\frac{2\left(1-q\right)}{2+\left(1-q\right)\left(n-k\right)}<1.
\label{eq:qp_q_r_case}
\end{equation}
In both (\ref{eq:qp_q_t_case}) and (\ref{eq:qp_q_r_case}), $q=1\Rightarrow q'=1$ in (\ref{eq:qp_q_r_case}), yielding the
classical property of Boltzmann systems, any subsystem of which is still of the Boltzmann type.

\vskip 3mm {\sf Case of convolution} \vskip 2mm \nd Choosing
$A=\left[a_{1},a_{2},\dots,a_{n}\right]$ in (\ref{eq:AKAT}) yields
the following results:

\[
X\sim\mathcal{T}\left(m,n,K\right)\Rightarrow\sum_{i=1}^{n}a_{i}X_{i}\sim\mathcal{T}\left(m,1,AKA^{T}\right)\]

\[
Y\sim\mathcal{R}\left(p,n,K\right)\Rightarrow\sum_{i=1}^{n}a_{i}Y_{i}\sim\mathcal{R}\left(p,1,AKA^{T}\right)\]
We note again that this stability result requires a special type of dependence
between the components $\left\{ X_{i}\right\} $ or $\left\{
Y_{i}\right\} $ namely the fact that they are extracted from the same
system.

\section{The stability issue for independent vectors}

\nd Few results exist about the convolution of two independent
Student-t or Student-r vectors.  In Ref. \cite{Oliveira}, Oliveira et al.
remark that if $X_{1}$ and $X_{2}$ are independent and
$\mathcal{T}\left(m,1,\sigma\right)$ distributed, their sum\[
Z=X_{1}+X_{2}\] can be very accurately approximated as a
$\mathcal{T}\left(m',1,\sigma'\right)$ for some
$\left(m',\sigma'\right)$ depending on $\left(m,\sigma\right).$
However, they provide only an approximation to the map
$\left(m,\sigma\right)\rightarrow\left(m',\sigma'\right).$

\nd An important result can be stated when $q>1,$ in the
special case for which the number $m$ of degrees of freedom is an odd integer $m=2l+1.$

\begin{theorem}\cite{berg}
If $X_{1}$ and $X_{2}$ are two independent vectors following
a distribution $\mathcal{T}\left(2l+1,n,\frac{I_{n}}{2l+1}\right)$ and
if $\alpha$ is such that $0\le\alpha\le1$ and $\beta=1-\alpha$,
then the distribution of\[
Z=\alpha X_{1}+\beta X_{2}\]
can be expressed as 
\begin{equation}
f_{Z}\left(Z\right)=\sum_{k=l}^{2l}\gamma_{k}^{\left(2l+1\right)}\left(\alpha\right)\mathcal{T}\left(2k+1,n,\frac{I_{n}}{2k+1}\right)
\label{eq:fz}\end{equation}
with
\begin{eqnarray*}
\gamma_{k}^{\left(l\right)}\left(\alpha\right) & = & \left(4\alpha\left(1-\alpha\right)\right)^{k}\left(\frac{l!}{\left(2l\right)!}\right)^{2}2^{-2l}\frac{\left(2l-2k\right)!\left(2l+2k\right)!}{\left(l-k\right)!\left(l+k\right)!}\\
 &  & \times\sum_{j=0}^{l-k} {2l+1 \choose 2j}{l-j \choose k}\left(2\alpha-1\right)^{2j},Ê\,\,\, 0\le k\le l\
\end{eqnarray*}
\end{theorem}
Since coefficients $\gamma_{k}^{(l)}$ are positive and sum to $1$ (see \cite{berg} for a proof), this result can be interpreted as follows: the convolution of
$\mathcal{T}$ distributions with odd degrees of freedom follows a
$\mathcal{T}$ distribution whose degrees of freedom are
randomized: 
\be
f_{Z}\left(Z\right)=\mathcal{T}\left(2l+2K+1,n,\frac{I_{n}}{2l+2K+1}\right)
\ee
where $K$ is a random variable defined as\[
\Pr\left\{ K=k\right\} =\gamma_{k}^{\left(l\right)}\left(\alpha\right), 0\le k\le l\]

\nd As an example, if $n=1$ and $m=3\Rightarrow l=1$, we have\[
f_{Z}\left(z\right)=\gamma_{1}^{\left(3\right)}\mathcal{T}\left(3,1,1/3\right)+\gamma_{2}^{\left(3\right)}\mathcal{T}\left(5,1,1/5\right)\]
with \[
\gamma_{1}^{\left(3\right)}=1-3\alpha\left(1-\alpha\right), \gamma_{2}^{\left(3\right)}=3\alpha\left(1-\alpha\right).\]
We note that conditions $0\le\alpha\le1$ and $\alpha+\beta=1$ are
not restrictive since

\begin{itemize}
\item if $\alpha<0$, then by parity of $\mathcal{T}\left(n,m,K\right)$,
$\alpha Y_{1}\sim\left(-\alpha\right)Y_{1}$
\item if $\alpha+\beta\ne1$, then $\alpha Y_{1}+\beta Y_{2}\sim\left(\alpha+\beta\right)\left(\frac{\alpha}{\alpha+\beta}Y_{1}+\frac{\beta}{\alpha+\beta}Y_{2}\right).$
\end{itemize}

An important result is the following one:
formula (\ref{eq:fz}) can be extended to the case where
$X_{1}\sim\mathcal{T}\left(m,n,K\right)$ and
$X_{2}\sim\mathcal{T}\left(m,n,K\right)$ provided $X_{1}$ and
$X_{2}$ have the same $q-$covariance matrix $K$: in that case,
$K^{-1/2}X_{1}$ and $K^{-1/2}X_{2}$ have identity $q-$covariance,
so that the distribution of $K^{-1/2}\left(X_{1}+X_{2}\right)$ can
be computed using formula (\ref{eq:fz}) and distribution of
$X_{1}+X_{2}$ can be obtained by a simple change of variable.

\section{Another approach to the stability issue: random convolution}

\nd A radically different approach to the problem we are
discussing here, namely,  the conditions of stability for
power-type distributions,  can be followed in the case $q<1$ by
considering the polar factorization property of the stochastic
representation (\ref{eq:stochastic_r}).

\begin{theorem}
If $\,Y$ has stochastic representation \[
Y=\frac{\Sigma^{1/2}G}{\sqrt{G^{T}G+b^{2}}}\] where $G$ is a
Gaussian vector with unit covariance matrix and $b$ is $\chi$
distributed, independent of $G$, then $Y$ is independent of
$\sqrt{G^{T}G+b^{2}}$; we remark that the later is chi distributed
with $p+2=\frac{2}{1-q}+n+2$ degrees of freedom.
\end{theorem}
\nd An important consequence of this property is that it allows to
derive a new kind of convolution of random type, as expressed by
the next theorem \cite{johnson}.

\begin{theorem}
If $Y_{1}\sim\mathcal{R}\left(p,n,K_{1}\right)$ and $Y_{2}\sim\mathcal{R}\left(p,n,K_{2}\right)$
are two independent vectors, if $\alpha_{1}$
and $\alpha_{2}$ are two real scalars and $a_{1}$ and $a_{2}$ are
two independent chi random variables with $d=p+2=\frac{2}{1-q}+n+2$
degrees of freedom, and if $\beta_{1}=\frac{a_{1}}{\sqrt{p+2}}\alpha_{1}, \beta_{2}=\frac{a_{2}}{\sqrt{p+2}}\alpha_{2}$
then vector\[
Y=\beta_{1}Y_{1}+\beta_{2}Y_{2}\]
is Gaussian with $q-$covariance matrix $R=\frac{p-n+2}{p+2}\left(\alpha_{1}^{2}K_{1}+\alpha_{2}^{2}K_{2}\right)$.
Moreover, if $c$ is chi distributed with $p-n+2$ degrees of freedom
and independent of $Y,$ then\[
Z=\frac{Y}{\sqrt{Y^{T}R^{-1}Y+c^{2}}}=\frac{\beta_{1}Y_{1}+\beta_{2}Y_{2}}{\sqrt{\left(\beta_{1}Y_{1}+\beta_{2}Y_{2}\right)^{T}R^{-1}\left(\beta_{1}Y_{1}+\beta_{2}Y_{2}\right)+c^{2}}}\]
 is again $\mathcal{R}\left(p,n,K\right)$ distributed with $q-$covariance matrix
$K=\frac{1}{p+2}\left(\alpha_{1}^{2}K_{1}+\alpha_{2}^{2}K_{2}\right).$
\end{theorem}
\nd In Figure \ref{fig4} below, the distribution of
$\beta_{1}=\alpha_{1}\frac{a_{1}}{\sqrt{p+2}}$ is represented for
$p+2=10, 20$ and $50,$ and $\alpha_{1}=2$.

\begin{figure}[h] 
   \centering
   \includegraphics[scale=0.4]{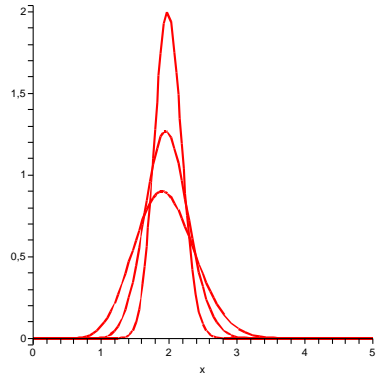}
   \caption{distribution of $\beta_{1}=\alpha_{1}\frac{a_{1}}{\sqrt{p+2}}$ for
$p+2=10, 20$ and $50,$ and $\alpha_{1}=2$}
   \label{fig4}
\end{figure}

\nd It is clearly seen that $\beta_{1}$ is a
"fluctuating'' version of the deterministic value
$\alpha_{1}=2$; since \be E\beta_{1}=\alpha_{1}
\sqrt{\frac{2}{p+2}}
\frac{\Gamma\left(\frac{p+3}{2}\right)}{\Gamma\left(\frac{p+2}{2}\right)},\ee
we have $\lim_{p\rightarrow+\infty}E\beta_{1}=\alpha_{1}$;
moreover, the variance of $\beta_{1}$ is 
\be var\left(\beta_{1}\right)=
\alpha_{1}^{2}\left(1-\frac{2}{p+2} \frac{\Gamma^{2}
\left(\frac{p+3}{2}\right)}{\Gamma^{2}\left(\frac{p+2}{2}\right)}\right),
\ee
so that $\lim_{p\rightarrow+\infty}var\left(\beta_{1}\right)=0$:
thus the number of degrees of freedom $p+2$ - imposed by the value
of $q$  that characterizes $Y_{1}$ and $Y_{2}$ through $p=n+\frac{2}{1-q}$ - rules the
fluctuation intensity  of $\beta_{1}$ around the deterministic value
$\alpha_{1}.$

\section{The Extensivity Issue}\label{sectionext}

\nd Still a different and important  question was raised  in
\cite{Tsallis}, namely  the extensivity issue: assuming that a
system $A$ is composed of two \textit{independent} subsystems $A_{1}$ and
$A_{2}$, the total $q-$entropy 
\[
S_{q}\left(A\right)=S_{q}\left(A_{1}\times
A_{2}\right)=S_{q}\left(A_{1}\right)+S_{q}
\left(A_{2}\right)+\left(1-q\right)S_{q}\left(A_{1}\right)S_{q}\left(A_{2}\right)
\]
is nonextensive  (i.e. nonadditive) unless $q=1,$ which characterizes
the Shannon entropy \footnote{this paragraph concerns only Tsallis entropy $S_{q}$ since the Harvda-Charvat-R\'{e}nyi entropy $H_{q}$ is extensive}. A natural question arises then: what kind
of dependence should exist between subsystems $A_{1}$ and $A_{2}$
so that $S_q$  becomes extensive ?

\nd An answer has been given to this question in the case of
Gaussian systems, as follows \cite{correlated}.

\begin{theorem}
If $0\le Q\le1$ and $n\in\mathbb{N}$ then there exists a positive
definite matrix $K$ and an $n-$variate Gaussian vector $X$ with
covariance matrix $K$ such that $X$ verifies the extensivity condition\begin{equation}
S_{Q}\left(X\right)=\sum_{i=1}^{n}S_{Q}\left(X_{i}\right).\label{eq:Sqext}\end{equation}

\end{theorem}

Trying to extend this
result to the distributions (\ref{eq:Student-t}) and
(\ref{eq:Student-r}), one should be careful about the following fact:
if $X$ is an $n-$variate random vector with probability
density (\ref{eq:Student-t}) or (\ref{eq:Student-r}) and non-extensivity parameter $q$
then any single component, say $X_{1}$, of $X$ is again
of the power type, but with a {\it different} nonextensivity
parameter, say $q_{1}$, related to $q$ via
(\ref{eq:qp_q_t_case}) or equivalently  (\ref{eq:qp_q_r_case}):
\[
q_{1}=1+\frac{2(q-1)}{2-\left(q-1\right)\left(n-1\right)}
\]

\nd Thus, the choice of $Q$ as related to $q$ and $q_{1}$ should
be decided. The choice $Q=2-q$ has a long history in the nonextensive
literature and already appeared in the paper \cite{curado} - for a
thorough discussion of the issue and its physical interpretation
see \cite{ferri}. This choice yields the following result.

\begin{theorem}
$\forall m>1$ and $n\in\mathbb{N},$ there exists a positive definite
matrix $K$ and an $n-$variate Student-t vector $X$ with $m$ degrees
of freedom and $q-$covariance matrix $K$ such that\[
H_{Q_{n}}\left(X\right)=\sum_{i=1}^{n}H_{Q_{1}}\left(X_{i}\right)\]
with $q_{1}=\frac{2q+\left(1-q\right)\left(n-1\right)}{2+\left(1-q\right)\left(n-1\right)}$,
$Q_{n}=2-q$ and $Q_{1}=2-q_{1}.$
\end{theorem}
\nd This result can be extended to the case $q<1$ as follows.

\begin{theorem}
$\forall p>1$ and $n\in\mathbb{N}$, there exists a positive definite
matrix $K$ and an $n-$variate Student-r vector $Y$ with $p$ degrees
of freedom and $q-$covariance matrix $K$ such that\[
H_{Q_{n}}\left(Y\right)=\sum_{i=1}^{n}H_{Q_{1}}\left(Y_{i}\right)\]
with $q_{1}=\frac{2+\left(n-3\right)\left(1-q\right)}{2+\left(n-1\right)\left(1-q\right)},$
$Q_{n}=2-q$ and $Q_{1}=2-q_{1}.$
\end{theorem}

\section{Conclusion}

\nd In this communication we have presented several results
concerning (i) the stability and (ii)
the extensivity of power-law random
vectors.We have shown that a certain kind of dependence between
the components of these vectors, namely the fact that they belong
to a larger system that is itself distributed \`a la power-law,
ensures stability of these variables. This property is a direct
consequence of the elliptical invariance of the associated $S_q$ or $H_{q}$ entropy.

\nd In the case of independent components, we have introduced a
random-type convolution that ensures stability for the power law
distributions.

\nd Finally, we have shown that $S_q$ can be additive if a
proper kind of correlation is introduced between the components of
the pertinent system, whose properties are to be described by
power-law vectors. Further work in progress concerns the extension
of this last result to the larger family of elliptically invariant
distributions.

\ack
C. V. wishes to thank Pr. Max-Olivier Hongler from EPFL for his
kind invitation to the Laboratoire de Production Microtechnique at E.P.F.L.

\end{document}